\documentclass[journal = jacsat]{achemso} 
\setkeys{acs}{articletitle=true}
\usepackage{graphicx}
\usepackage{lscape}
\usepackage{booktabs}
\usepackage{color}
\usepackage{bm}
\usepackage{caption}
\usepackage{achemso}
\usepackage{amssymb,amsmath}
\usepackage[sort&compress,numbers,super]{natbib}
\title{Electrostatic forces from reactive molecular orbitals driving chemical reactions}
\author{Takao Tsuneda}
\affiliation{Department of Chemistry, Faculty of Science, Hokkaido University, Sapporo 060-0810, Japan}
\alsoaffiliation{Graduate School of System Informatics, Kobe University, Nada-ku, Kobe, Hyogo 657-8501, Japan}
\email{takaotsuneda@sci.hokudai.ac.jp}
\author{Tetsuya Taketsugu}
\affiliation{Department of Chemistry, Faculty of Science, Hokkaido University, Sapporo 060-0810, Japan}
\alsoaffiliation{Institute for Chemical Reaction Design and Discovery (WPI-ICReDD), Hokkaido University, Sapporo 001-0021, Japan}

\begin{document}


\begin{abstract}
This study offers a physics-based framework for understanding chemical reactions,
unveiling the pivotal role of the occupied reactive orbital (ORO), 
the most stabilized occupied molecular orbital during a reaction,
in driving atomic nuclei along the reaction pathway via electrostatic forces.
We show that these electrostatic forces are governed by the negative gradient
of orbital energy, establishing a direct link between molecular orbital energy variations
and nuclear motion.
The forces generated by OROs, termed reactive-orbital-based electrostatic forces (ROEFs),
were systematically analyzed across 48 representative reactions.
Our findings reveal that reactions can be classified into four distinct types,
with two dominant types emerging: those that maintain reaction-direction ROEFs
either from the early stages or immediately preceding the transition state.
These ROEFs carve distinct grooves along the intrinsic reaction coordinates
on the potential energy surface, shaping the reaction pathway.
Notably, ORO variations align directly with the curly arrow diagrams widely employed
in organic chemistry, bridging the curly arrow-like representation of electron transfer
with the rigorous potential energy surface framework.
This connection highlights the integration of electronic and nuclear motion theories,
offering a unified perspective on the forces that drive chemical transformations.
By linking orbital energy variations to nuclear motions,
this study establishes a robust framework for understanding the interplay
between electronic structure and reaction mechanisms.
\end{abstract}

\maketitle





What drives chemical reactions: electron motion or nuclear motion?
The theoretical understanding of chemical reactions, grounded in the elucidation
of molecular electronic structures via quantum mechanics, has historically diverged
into two perspectives: electronic theories, such as the theory of organic reaction
mechanisms \cite{grossman2019} and frontier orbital theory \cite{fukui1952},
which emphasize electron motion, and nuclear motion theories,
rooted in the potential energy surface (PES) framework \cite{schlegel2003},
which focus on atomic nuclei motions.
Despite addressing the same fundamental phenomenon, the interrelation between
these theories remains largely unexplored.
Electronic theories propose that electron motion orchestrates molecular structural
transformations during chemical reactions.
However, this assertion has neither been rigorously validated nor supported
by quantitative evidence.
Conversely, nuclear motion theories, which quantitatively describe energetic changes
related to nuclear motions, have become the dominant paradigm for predicting
reaction rates.
But is it truly impossible to bridge these perspectives?

To reconcile electron and nuclear motions in chemical reactions,
it is essential to identify the specific electron motions that dictate reaction pathways.
Reactive orbital energy theory (ROET) \cite{tsuneda-singh2014} addresses
this challenge by leveraging a statistical mechanical framework \cite{tsuneda_etal2016}
to identify the molecular orbitals, both occupied and unoccupied,
with the largest orbital energy variations before and after the reaction as the reactive orbitals.
Interestingly, reactive orbitals identified by ROET are often neither the HOMO nor the LUMO.
This distinction becomes particularly pronounced in catalytic reactions
involving transition metals, where low-energy valence orbitals with high electron densities
frequently serve as the reactive orbitals.
The development of ROET was made possible by advancements in
long-range corrected (LC) density functional theory (DFT) \cite{tsuneda-hirao2014, tsuneda_etal2010},
enabling accurate and quantitative orbital energy calculations.
Recent comparative studies of molecular orbital densities obtained from LC-DFT canonical orbitals
and Dyson orbitals derived from the coupled-cluster method have demonstrated
the exceptional fidelity of LC-DFT in replicating both orbital shapes and energies
\cite{hasebe_etal2023}.
ROET analysis of organic reactions revealed that transitions between reactive orbitals
correspond closely to the directions of curly arrows used to represent reaction mechanisms
\cite{tsuneda_etal2023}.
Furthermore, ROET applied to the comprehensive reaction pathways of glycine demonstrated
a one-to-one correspondence between reaction pathways and their respective reactive orbitals,
offering a novel perspective on electron motion in chemical reactions \cite{hasebe_etal2021}.

To connect these electron motions with nuclear motions, we turn to electrostatic force
theory \cite{nakatsuji1973}, which quantifies the forces exerted by electronic configurations
on molecular nuclei through Hellmann-Feynman forces \cite{feynman1939}.
By integrating ROET with electrostatic force theory, it becomes possible to determine
the forces exerted by reaction-driving electrons on nuclei and to evaluate
their alignment with the reaction pathway.
When these forces align with the reaction direction, they carve reaction pathways on the PES,
directly linking electron motion to nuclear motion.

In this study, we elucidate the driving forces of chemical reactions by calculating
the electrostatic forces exerted by reaction-driving electrons on the nuclei of molecules.
Using reactive orbitals identified through ROET, we compute the Hellmann-Feynman forces
within the framework of electrostatic force theory.
This integrated approach allows us to rigorously test the long-standing assumption
in electronic theories that electron motion directs molecular structural transformations,
providing a unified framework for understanding the interplay between electron motions
and nuclear motions in chemical reactions.


\section*{Theory \label{sec:theory}}

\subsection*{Electrostatic forces exerted by electrons on nuclei}

To understand the forces acting on nuclei during chemical reactions,
we begin by examining the electrostatic forces exerted by all electrons
on the nuclei within a molecule.
The Hamiltonian, $\hat{H}$, governing the system of electrons and nuclei \cite{tsuneda2014},
is given by
\begin{eqnarray}
\hat{H} = \sum_i^{n_{\rm elec}} \left( -\frac{1}{2}\nabla_i^2 - \sum_A^{n_{\rm nuc}} \frac{Z_A}{r_{iA}} \right)
+ \sum_{i < j}^{n_{\rm elec}} \frac{1}{r_{ij}} 
+ \sum_{A < B}^{n_{\rm nuc}} \frac{Z_A Z_B}{R_{AB}},
\label{eqn:hamiltonian}
\end{eqnarray}
where $\nabla_i^2$ denotes the Laplacian with respect to electron $i$,
$r_{iA}$ is the distance from electron $i$ to nucleus $A$,
$r_{ij}$ is the inter-electronic distance between electrons $i$ and $j$,
$R_{AB}$ represents the internuclear distance between nuclei $A$ and $B$,
$Z_A$ is the charge of nucleus $A$, and
$n_{\rm elec}$ and $n_{\rm nuc}$ represent the numbers of electrons and nuclei, respectively.
Atomic units are used ($\hbar = e^{2} = m = 1$, energies are in hartree,
and distances are in bohr).
Building upon electrostatic theory \cite{nakatsuji1973},
the Hellmann-Feynman force exerted by the electrons and nuclei on nucleus $A$ is expressed as
\begin{eqnarray}
{\bf F}_A &=& 
- \frac{\partial}{\partial {\bf R}_A} 
\Big\langle \Psi \left| \hat{H} \right| \Psi \Big\rangle 
= -\Bigg\langle \Psi \left| \frac{\partial \hat{H}}{\partial {\bf R}_A} \right| \Psi \Bigg\rangle
\nonumber \\
&=& Z_A \int d{\bf r} \rho({\bf r}) \frac{{\bf r}-{\bf R}_{A}}
{|{\bf r}-{\bf R}_{A}|^3}
- Z_A \sum_{B(\neq A)}^{n_{\rm nuc}} Z_B \frac{{\bf R}_{AB}}{{R_{AB}}^3} ,
\label{eqn:electrostatic_force}
\end{eqnarray}
where $\rho(\mathbf{r})$ represents the electron density at position $\mathbf{r}$,
and $\mathbf{R}_{A}$ and $\mathbf{R}_{AB}$ are position vector of
nucleus $A$ and vector from nucleus $A$ to nucleus $B$, respectively.
According to the Hellmann-Feynman theorem, these forces represent classical electrostatic forces
when the electron distribution is determined variationally \cite{feynman1939}.
For a wavefunction describing all electrons, the electron density is given by
\begin{eqnarray}
\rho({\bf r}_1) = N \int ds_1 d{\bf x}_2 \cdots d{\bf x}_N
\Psi^*({\bf x}_1,{\bf x}_2,\cdots,{\bf x}_N) \times
\Psi({\bf x}_1,{\bf x}_2,\cdots,{\bf x}_N),
\label{eqn:density_exact}
\end{eqnarray}
which is integrated over all spatial and spin coordinates,
${\bf x}_n = ({\bf r}_n,s_n)$ for $n=$ 1 to $N$, except $\mathbf{r}_1$.
Within independent electron approximations \cite{hartree1928}, such as
the Kohn-Sham method \cite{kohn-sham1965}, the electron density simplifies to
\begin{eqnarray}
\rho({\bf r}) = \sum_i^{n_{\rm elec}} \rho_i({\bf r}) 
= \sum_i^{n_{\rm elec}} \phi_i^* ({\bf r}) \phi_i ({\bf r}),
\label{eqn:density_independent-electron-model}
\end{eqnarray}
where $\phi_i$ is the $i$-th spin orbital wavefunction.
The total electrostatic force on nucleus $A$ is then expressed as the sum of
contributions from electrons (${\bf F}_A^{\rm elec}$) and other nuclei (${\bf F}_A^{\rm nuc}$),
\begin{eqnarray}
{\bf F}_A &=& {\bf F}_A^{\rm elec} +  {\bf F}_A^{\rm nuc}
= Z_A \sum_i^{n_{\rm elec}} {\bf f}_{iA} 
- Z_A \sum_{B(\neq A)}^{n_{\rm nuc}} Z_B \frac{{\bf R}_{AB}}{{R_{AB}}^3},
\label{eqn:electrostatic_force_ipm} \\
\end{eqnarray}
where the force contribution from the $i$-th orbital on nucleus $A$ is
\begin{eqnarray}
{\bf f}_{iA} &=& \int d{\bf r} \phi_i^*({\bf r}) 
\frac{{\bf r}-{\bf R}_{A}}{|{\bf r}-{\bf R}_{A}|^3} \phi_i({\bf r}),
\label{eqn:electrostatic_force_ipm2} 
\end{eqnarray}
Using this framework, the influence of reactive orbitals on nuclear forces can be isolated.
The force contribution from an occupied reactive orbital (ORO) is given by
\begin{eqnarray}
{\bf f}_{A}^{\rm ORO}
&=& \int d{\bf r} \phi^{\rm ORO*}({\bf r})
\frac{{\bf r}-{\bf R}_{A}}{|{\bf r}-{\bf R}_{A}|^3} \phi^{\rm ORO} ({\bf r}),
\label{eqn:electrostatic_force_roet} 
\end{eqnarray}
where $\phi^{\rm ORO}$ represents the wavefunction of the ORO.
This formulation allows for the direct assessment of how variations in reactive orbitals
influence nuclear forces during chemical reactions.

In this study, we investigate the primary electrostatic forces driving nuclear motions
in chemical reactions, focusing specifically on the forces generated by variations in the ORO.
The progression of electron transfer throughout a reaction is characterized by changes
in OROs \cite{tsuneda_etal2023}.
The electrostatic force vector resulting from ORO variations can be expressed as:
\begin{eqnarray}
{\bf F}_A^{\rm ROEF} 
= Z_A {\bf f}_{A}^{\rm ORO}.
\label{eqn:roef_occ-variance}
\end{eqnarray}
where we refer to ${\bf F}_A^{\rm ROEF}$ as the reactive orbital-based electrostatic force (ROEF)
vector.
This formulation isolates the contribution of ORO variations along the reaction pathway,
which may involve interactions with unoccupied orbitals, while neglecting the contributions
from other orbitals.

\subsection*{Relationship between electrostatic forces and orbital energies}

Next, we examine the relationship between ROEF and the Kohn-Sham orbital energies \cite{tsuneda2014}.
The Kohn-Sham orbital energy, $\epsilon_i$, is defined as
\begin{eqnarray}
\epsilon_i &=& h_i + \sum_{j}^{n_{\rm elec}} J_{ij} + \int d{\bf r} \rho_i({\bf r}) v_{\rm xc},
\label{eqn:electrostatic_force_orbital4} 
\end{eqnarray}
where $v_{\rm xc}$ is an exchange-correlation potential functional,
and $h_i$ represents the one-electron Hamiltonian, given by
\begin{eqnarray}
h_i &=& \int d{\bf r} 
\phi_i^*({\bf r}) \left\{ -\frac{1}{2}\nabla^2-\sum_A^{n_{\rm nuc}} \frac{Z_A}{r_{iA}}
\right\} \phi_i({\bf r}).
\label{eqn:electrostatic_force_orbital2} 
\end{eqnarray}
The derivative of the orbital energy with respect to the coordinates of nucleus $A$
is then expressed as
\begin{eqnarray}
\frac{\partial \epsilon_i}{\partial {\bf R}_A} 
= \frac{\partial}{\partial {\bf R}_A} 
\left\{ 
h_i + \sum_{j}^{n_{\rm elec}} J_{ij} + \int d{\bf r} \rho_i({\bf r}) v_{\rm xc}
\right\}
= \frac{\partial h_i}{\partial {\bf R}_A}.
\label{eqn:electrostatic_force_orbital5} 
\end{eqnarray}
where contributions from the $v_{\rm xc}$ term are negligible due to its limited
explicit dependence on nuclear positions.

\subsection*{Reactive orbital-based electrostatic force (ROEF)}

The total electrostatic force vector on nucleus $A$, derived from the Kohn-Sham electronic energy,
$E_{\rm KS}$, can be written as
\begin{eqnarray}
{\bf F}_A^{\rm elec} &=& - \frac{\partial E_{\rm KS}}{\partial {\bf R}_A}
= - \frac{\partial}{\partial {\bf R}_A} \left\{
\sum_i^{n_{\rm elec}} h_i + \sum_{i}^{n_{\rm elec}}\sum_{j(\neq i)}^{n_{\rm elec}} J_{ij}
+ E_{\rm xc} \right\} \nonumber \\
&=& - \sum_i^{n_{\rm elec}} \frac{\partial h_i}{\partial {\bf R}_A}
= - \sum_i^{n_{\rm elec}} \frac{\partial \epsilon_i}{\partial {\bf R}_A},
\label{eqn:electrostatic_force_orbital5.5} 
\end{eqnarray}
indicating that the force contribution from each orbital is proportional
to its energy gradient with respect to nuclear coordinates.
Substituting this relationship, the ROEF vector can be expressed as
\begin{eqnarray}
{\bf F}_A^{\rm ROEF} &=& -\frac{\partial \epsilon^{\rm ORO}}{\partial {\bf R}_A},
\label{eqn:electrostatic_force_orbital6} 
\end{eqnarray}
where $\epsilon^{\rm ORO}$ is the orbital energy of the ORO.
This formulation highlights that the electrostatic force vector exerted by an orbital
is determined by the gradient of its energy with respect to the nuclear coordinates.
By focusing on the ORO, this approach establishes a direct connection between
orbital energy variations and nuclear forces, enabling a more detailed understanding
of the forces driving chemical reactions at the atomic level.

\subsection*{Accuracy and interpretation of ROEFs}

From the relationship established in Eq. (\ref{eqn:electrostatic_force_orbital6}),
we infer that orbital energy variations influence electrostatic forces
during chemical reactions.
Specifically, for OROs that exhibit decreasing orbital energies as the reaction progresses,
the resulting electrostatic forces are expected to align with the reaction direction,
as described in Eq. (\ref{eqn:roef_occ-variance}).
These reaction-aligned electrostatic forces create trenches on the PES,
delineating the intrinsic reaction coordinate (IRC) and defining the reaction pathway.
OROs act as the driving force for a reaction when their electrostatic forces are sufficiently
strong to establish the IRC.
Since orbital energy remains constant during idealized electron transfer \cite{sham-schluter1985},
orbital energy variations are minimal in reaction stages dominated by electron transfer.
This behavior is reflected in the early stages of many reactions,
as shown in Figs. S1 and S2 of the Supporting Information.
According to Eq. (\ref{eqn:electrostatic_force_orbital6}), when the orbital energy gradient
approaches zero, the corresponding electrostatic force vector vanishes.
However, even small gradients in orbital energy, arising from structural transformations
along the IRC, determine the magnitude of the electrostatic force.
Accurate calculation of these forces, therefore, hinges on precise determination
of orbital energy variations induced by structural changes.
This underscores the necessity of a robust theoretical framework capable
of reliably predicting orbital energy gradients across different molecular structures.
LC-DFT provides a critical tool in this regard, as it quantitatively reproduces orbital energies
with high fidelity, making it indispensable for such analyses.

The Pulay force, an artificial force arising from the use of basis sets,
is a factor potentially related to ROEFs \cite{pulay1970, nakatsuji_etal1982, bakken_etal1982}.
Previous studies report that the magnitude of Pulay forces for all electrons
is less than 10 kJ mol$^{-1}$ bohr$^{-1}$, with even smaller contributions
for individual molecular orbitals.
By contrast, ROEFs derived from orbital energy gradients are on the order of several hundred
kJ mol$^{-1}$ bohr$^{-1}$.
This significant disparity indicates that Pulay forces have negligible influence on ROEFs,
and they are therefore excluded from consideration in this study.

An important insight from the interplay between ROEFs and orbital energy variations
relates to the virial theorem \cite{rodriguez_etal2009},
which states that the kinetic energy is half the potential energy but with the opposite sign
for nuclear-electron interactions.
Under independent electron approximations, this relationship holds for individual electrons
\cite{harbola1998}.
When orbital energy variations during orbital mixing are small,
nuclear-electron potential changes are similarly minimized,
suppressing fluctuations in ROEFs, as shown in Eqs. (\ref{eqn:electrostatic_force_orbital2})
and (\ref{eqn:electrostatic_force_orbital5.5}).
While electron transfer induces polarization in the electron distribution,
ROEFs remain stable.
However, structural deformations near the transition state can generate significant ROEFs,
influencing the reaction pathway.

This study highlights the critical role of electron transfer as the driving force
for structural transformations during reactions.
By calculating the electrostatic forces associated with ORO energy gradients
along the IRC, we trace the forces exerted on atomic nuclei,
revealing how changes in electron distribution induced by electron transfer shape
the reaction pathway.


\subsection*{Three-body model for evaluating ROEFs}

Using the framework of ROEFs, we analyze a diverse set of atom-transfer reactions,
comprising 16 hydrogen transfer reactions, one heavy atom transfer,
two nucleophilic substitutions, and five unimolecular reactions.
Transition state and IRC calculations were performed with the LC-BLYP+LRD/aug-cc-pVTZ method
\cite{tsuneda_etal2001a, sato-nakai2009} using the GAMESS program \cite{gamess}.
Molecular orbital correspondences along the IRCs were established with
an in-house program designed to automatically trace orbital energies
and wavefunctions throughout the reaction pathway \cite{tsuneda_etal2023}.
The reaction pathway with the higher experimental reaction rate was designated
as the forward process, while the reverse process was treated as the backward process.
The ORO was identified as the orbital maximizing the relative change in orbital energy
between reactants and products. This change is quantified as
$2(\epsilon_i^{\rm prod}-\epsilon_i^{\rm reac})/
|\epsilon_i^{\rm prod}+\epsilon_i^{\rm reac}|$,
where $\epsilon_i^{\rm reac}$ and $\epsilon_i^{\rm prod}$ are the energies of the 
$i$-th molecular orbital in the reactant and product states, respectively.
For the ROEF analysis, atom-transfer reactions were modeled using a three-body approach.
\begin{figure}
\begin{center}
\includegraphics[width=17cm]{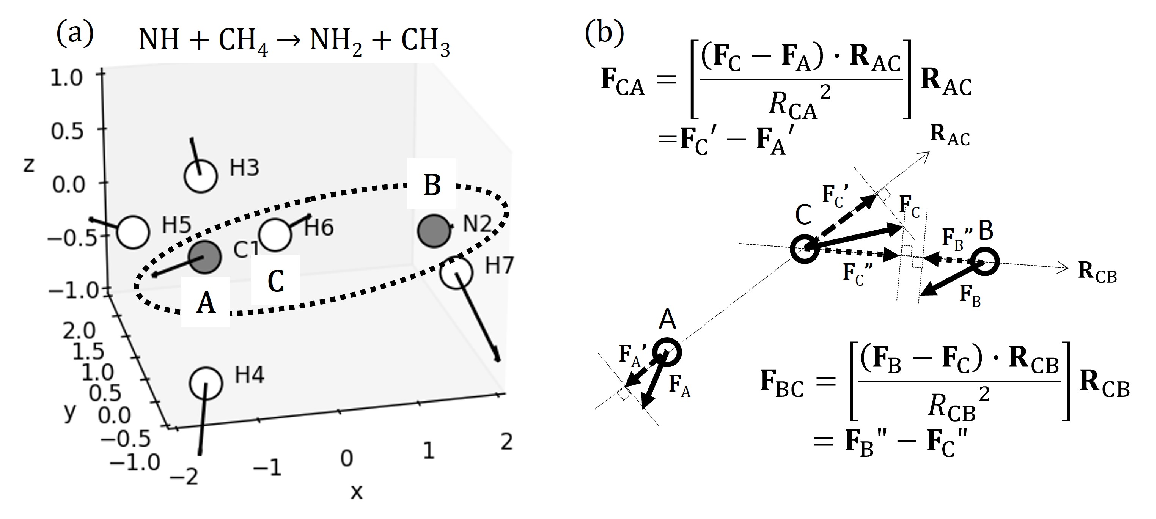}
\end{center}
\caption{
Three-body model for evaluating ROEFs in atom-transfer reactions:
(a) Example calculation model for the NH + CH$_4$ $\rightarrow$ NH$_2$ + CH$_3$ reaction
at the transition state,
where atoms C1, N2, and H6 are labeled as A, B, and C, and the movement of
atom C from A to B is considered.
(b) Method for calculating ROEFs in atom-transfer reactions:
${\bf F}_{\rm A}$, ${\bf F}_{\rm B}$, and ${\bf F}_{\rm C}$ are
the ROEF vectors calculated for atoms A, B, and C respectively,
${\bf F}_{\rm A}'$ and ${\bf F}_{\rm C}'$ are the projection vectors
of ${\bf F}_{\rm A}$ and ${\bf F}_{\rm C}$ in the direction of the A-C bond,
${\bf F}_{\rm C}"$ and ${\bf F}_{\rm B}"$ are the projection vectors of
${\bf F}_{\rm C}$ and ${\bf F}_{\rm B}$ in the direction of the C-B bond.
These projections enable the calculation of ROEF vectors,
${\bf F}_{\rm BC}$ and ${\bf F}_{\rm CA}$, crucial to the reaction.
For unimolecular reactions, only one ROEF vector set corresponds
to the reaction direction, and the remaining vectors are set to zero.
\label{fig:roef_calculation_model}}
\end{figure}
Figure \ref{fig:roef_calculation_model} illustrates this framework, applied to
the the NH + CH$_4$ $\rightarrow$ NH$_2$ + CH$_3$ reaction (a) and
the ROEF vector calculation method (b).
These ROEF values provide insight into the forces driving atomic movement
caused by electron redistribution during ORO variations.
ROEF variations along the IRCs for all reactions, along with corresponding
molecular structures and atomic coordinates, are presented
in Figs. S1 and S2 of the Supporting Information.



\section*{Results and Discussions \label{sec:results_and_discussions}}

\subsection*{Reaction types classified by ROEFs}

\begin{figure}
\begin{center}
\includegraphics[width=17cm]{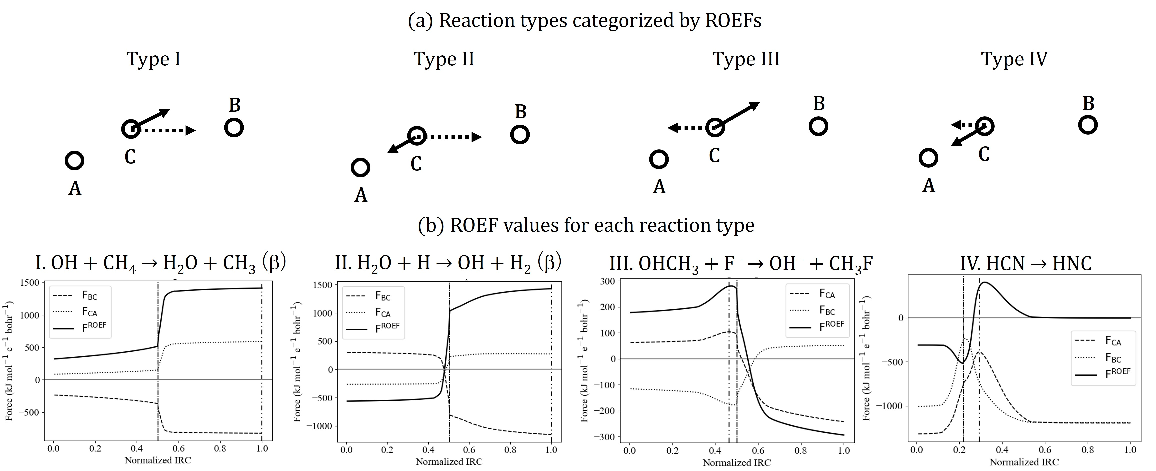}
\end{center}
\caption{
Variations in ROEF vectors categorized into four reaction types and their corresponding
ROEF changes: (a) conceptual diagrams of the three-body model depicting ROEF vectors
for four reaction types, with solid and dotted arrows representing electron transfer
before and after the TS, respectively;
(b) representative reactions for each type, illustrating ROEF variations along normalized IRCs.
The force magnitudes for the B-C and C-A bonds, $F_{\text{BC}}=|{\bf F}_{\rm BC}|$ and
$F_{\text{CA}}=|{\bf F}_{\rm CA}|$,
are shown as dotted and dashed curves, respectively, with the combined ROEF
($F_{\rm ROEF} = F_{\rm CA} - F_{\rm BC}$) depicted as a solid curve.
Definitions of $F_{\rm BC}$ and $F_{\rm CA}$ vectors are provided
in Fig. \ref{fig:roef_calculation_model}(b).
Dash-dot lines indicate TS locations, while dash-dot-dot lines mark points where
$F_{\rm ROEF}$ reaches its maximum.
For open-shell reactions, the spin states of the reactive orbitals contributing
the largest ROEF values are noted in parentheses next to the reaction formulas.
\label{fig:roef_variations_exam}}
\end{figure}

To elucidate the role of ROEFs in chemical reactions, we examined their behavior
along IRCs and identified four distinct patterns of ROEF variations.
Figure \ref{fig:roef_variations_exam}(a) provides a conceptual diagram of ROEFs
for each reaction type, using the three-body model shown
in Fig. \ref{fig:roef_calculation_model}(b).
ROEFs were classified into the following four patterns:
\begin{enumerate}
\renewcommand{\theenumi}{\Roman{enumi}}
\item
Continuous forward ROEFs throughout the reaction
\item
Sudden forward ROEFs just before the TSs, maintained afterward
\item
Forward ROEFs until the TSs, followed by backward ROEFs
\item
Backward ROEFs even after the TSs
\end{enumerate}
Figure \ref{fig:roef_variations_exam}(b) illustrates representative reactions
and ROEF variations for each type.
The force magnitudes $F_{\rm BC}$ and $F_{\rm CA}$, based on definitions
in Fig. \ref{fig:roef_calculation_model}(b), were calculated for the IRCs
of four characteristic atom-transfer reactions.
When $F_{\rm BC}$ is negative and $F_{\rm CA}$ is positive,
the electrostatic forces are aligned with the reaction direction.
The maximum force along the reaction path occurs where the difference
$F_{\rm CA} - F_{\rm BC}$ (solid curve) is greatest.
Figures S1 and S2 present ROEF variations for both forward and backward processes.
Forward ROEFs act along the reaction direction, while backward ROEFs oppose it.
The distribution of reactions across the four categories is summarized
in Fig. \ref{fig:roef_variations_exam}.
Forward reactions are predominantly type I (11 reactions) and type II (10 reactions),
with fewer in types III (2 reactions) and IV (1 reaction).
Similarly, backward reactions are primarily type I (12 reactions)
and type II (9 reactions), with only 1 and 2 reactions in types III and IV, respectively.
Most reactions exhibit forward ROEFs immediately before the TS, which continue beyond the TS.

\subsection*{Peak ROEF values for forward reaction processes}

\begin{figure}
\begin{center}
\includegraphics[width=16cm]{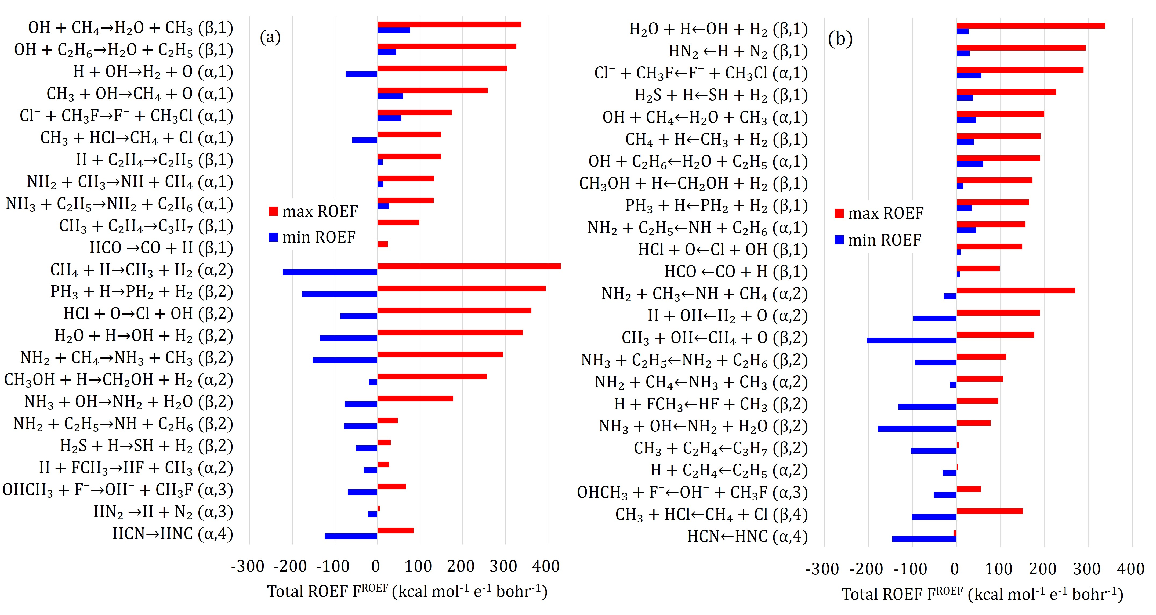}
\end{center}
\caption{
Peak ROEF values calculated from OROs for forward (a) and backward (b) processes
of 24 atom-transfer reactions:
black bars represent maximum forward ROEFs, while white bars show minimum backward ROEFs.
These ROEF values correspond to the combined force $F^{\rm ROEF} = F_{\rm CA} - F_{\rm BC}$
along the B-C and C-A bonds, as defined in Fig. \ref{fig:roef_calculation_model}(b).
Reaction formulas are accompanied by the spin states of the reactive orbitals
(noted in parentheses), along with their corresponding reaction types.
\label{fig:roef_occ_forward_backward}}
\end{figure}

Figure \ref{fig:roef_occ_forward_backward}(a) highlights the peak ROEF values
along the reaction pathways for forward processes, derived from ORO variations.
In many forward reactions, peak forward ROEFs significantly exceed the backward ROEFs.
Notably, higher backward ROEF peaks are observed only in three type II reactions,
two type III reactions, and one type IV reaction.
The substantial forward ROEFs in types I and II suggest strong forces acting
on atomic nuclei near and beyond the TS, implying that OROs generate trenches
along the IRCs on the PES for most forward processes.

Reactions classified as types I and II dominate forward processes,
accounting for 21 out of 24 reactions (detailed ROEF variations are provided in Fig. S1).
In type I reactions, ROEFs are consistently directed along the reaction path from the onset,
remaining constant throughout the reaction.
Examples include reactions involving hydrogen atom exchange with
OH, CH$_3$, or C$_2$H$_5$ molecules.
These reactions are primarily driven by an electrostatic potential gradient established
by the electron distribution from the start.
In type II reactions, ROEFs initially oppose the reaction progression but shift direction
just before the TS, aligning with the reaction path and persisting beyond the TS.
This behavior is frequently observed in reactions producing hydrogen or NH$_2$
molecules through hydrogen atom exchange.
The shift in ROEF direction occurs because orbital energy remains constant
during electron transfer, and accumulated energy is released through structural deformations
near the TS.
These reactions are electron transfer-driven, with forces acting primarily near the TS
or from the beginning of the reaction.
In both types I and II, ROEFs create depressions along the IRC on the PES,
forming the reaction pathway.
However, for type II reactions, this effect is more localized to the region near the TS.

\subsection*{Peak ROEF values for backward reaction processes}

Figure \ref{fig:roef_occ_forward_backward}(b) illustrates the peak directional ROEF values
derived from OROs for backward processes.
In type II reactions, five cases exhibit higher backward ROEFs compared to forward ones,
whereas this occurs in only one type IV reaction.
The elevated backward ROEFs in type II likely contribute to slower reaction rates.
In most other reactions, forward ROEFs dominate during the early stages,
generating electrostatic fields that carve trenches along the IRC on the PES.
Notably, electron transfer also contributes to shaping the IRC in backward processes.

Backward processes show a similar distribution to forward processes,
with 21 out of 24 reactions classified as types I and II (see Fig. S2).
This finding indicates that electron-driven forces influencing reaction progression
are generally applied near the TS or from the start.
A notable difference is the presence of an additional type IV reaction in the backward processes.
Type IV reactions exhibit backward ROEFs even beyond the TS,
indicating a persistent electron distribution bias in the backward direction.
This bias does not drive the reaction forward, explaining the slower rates
observed in these reactions.
For example, the CH$_4$ + Cl $\rightarrow$ CH$_3$ + HCl reaction
proceeds electrostatically without such a bias.
Conversely, the HNC $\rightarrow$ HCN reaction involves an H atom
circling the NC molecule, rendering the three-body model
in Fig. \ref{fig:roef_calculation_model}(b) less applicable.
Nevertheless, many backward processes remain influenced by electrostatic forces
arising from ORO electron distributions.

\subsection*{The roles of ROEFs as the driving force of chemical reactions}

These findings reveal that changes in the electron distribution within
OROs generate electrostatic forces acting on atomic nuclei,
playing a pivotal role in shaping the IRC on the PES.
Furthermore, they establish a direct connection between electronic motion theories
and nuclear motion theories, encompassing both forward and reverse reaction processes.
A particularly noteworthy insight is that, for each reaction pathway,
the electrostatic forces driving the reaction predominantly originate from the ORO
that most effectively lowers orbital energy in the reaction direction. 
In reverse reactions, analysis of ROEFs generated by the ORO that most significantly
reduces orbital energy in the reverse direction demonstrates that these forces naturally
act on the nuclei, driving the reaction backward.
Additionally, when orbital energy displays peak- or valley-like patterns,
reverse-directed electrostatic forces emerge before and after the extrema, respectively.
This behavior underscores the nuanced interplay between orbital energy variations
and nuclear forces, highlighting the critical role of OROs in guiding reaction dynamics.
These findings strongly support the conclusion that changes in the ORO that most effectively
lowers orbital energy in the reaction direction serve as the primary driving force
behind chemical reactions, uniting electronic and nuclear motion theories
under a cohesive framework.




In conclusion, this study elucidates the pivotal role of ROEFs,
arising from variations in OROs, in driving the motion of nuclei within reacting molecules.
These findings bridge the electron transfer processes described in curly arrow-like
representation with the concepts of the PES.
ROEFs, defined by the negative gradients of ORO energies, facilitate reaction progress
by mediating electron transfer.
Our analysis of 24 atom-transfer reactions, encompassing both forward and backward mechanisms,
revealed that ROEFs can be classified into four distinct types:
I. continuous forward ROEFs throughout the reaction,
II. sudden forward ROEFs just before the TSs that are maintained afterward,
III. forward ROEFs until the TSs, followed by backward ROEFs, and
IV. backward ROEFs even after the TSs.
Notably, reactions in types I and II dominate both forward and backward processes,
indicating that OROs primarily generate forward-directed ROEFs after the TS.
Furthermore, OROs produce significant positive ROEFs that directly drive reaction progression.
These observations underscore the critical role of ROEFs in shaping
the IRCs on the PES.

This study demonstrates that chemical reactions are driven
by electrostatic forces generated from ORO variations, which propel atomic nuclei
along the reaction pathway and define tracks on the PES.
Through these insights, we establish a meaningful connection between electron motion theories
and nuclear motion theories grounded in PES frameworks,
offering a unified perspective on the mechanics of chemical reactions.
Finally, it is noteworthy that ROEF- and ROET-based analyses rely solely on molecular orbitals
and orbital energies, making these approaches theoretically extendable
to {\it ab initio} wavefunction methods using Dyson orbitals \cite{dyson1949,ortiz2020}.




\section*{Acknowledgments}

This research was financially supported by JST CREST, Japan (Grant No. JPMJCR1902).

\section*{Supplementary material available}

The ROEF variations, stemming from variations in the OROs
across both forward and backward processes of all 24 atom-transfer reactions 
are detailed in Figs. S1 and S2, respectively.

\clearpage



\providecommand{\latin}[1]{#1}
\providecommand*\mcitethebibliography{\thebibliography}
\csname @ifundefined\endcsname{endmcitethebibliography}
  {\let\endmcitethebibliography\endthebibliography}{}

\end{document}